\documentclass[twocolumn,times]{aastex6}

\usepackage{natbib}

\newcommand{\kms}{km\,s$^{-1}$}
\newcommand{\cii}{[\ion{C}{2}]}
\newcommand{\ci}{[\ion{C}{1}]}

\newcommand{\lfir}{$L_{\mathrm{FIR}}$}

\newcommand{\lcii}{$L_\mathrm{[CII]}$}
\newcommand{\lcovi}{$L_{\mathrm{CO(6-5)}}$}
\newcommand{\lcovii}{$L_{\mathrm{CO(7-6)}}$}
\newcommand{\lpcovi}{$L^\prime_{\mathrm{CO(6-5)}}$}
\newcommand{\lpcovii}{$L^\prime_{\mathrm{CO(7-6)}}$}

\newcommand{\lsun}{$L_\sun$}
\newcommand{\msun}{$M_\sun$}

\newcommand{\msunyr}{$M_\sun$\,yr$^{-1}$}

\newcommand{\jykms}{Jy\,km\,s$^{-1}$}

\newcommand{\kkmspc}{K\,\kms\,pc$^2$}

\slugcomment{Accepted for publication}

\shorttitle{Molecular gas in three $z\sim7$ quasar hosts}
\shortauthors{Venemans et al.}

\begin{document}

\title{Molecular gas in three $z\sim7$ quasar host galaxies}

\author{Bram P.\ Venemans\altaffilmark{1},
Fabian Walter\altaffilmark{1},
Roberto Decarli\altaffilmark{1},
Carl Ferkinhoff\altaffilmark{2},
Axel Wei\ss\altaffilmark{3},
Joseph R.\ Findlay\altaffilmark{4},
Richard G.\ McMahon\altaffilmark{5,6},
Will J. Sutherland\altaffilmark{7},
Rowin Meijerink\altaffilmark{8}
}
\altaffiltext{1}{Max-Planck Institute for Astronomy, K{\"o}nigstuhl 17, 69117
  Heidelberg, Germany}
\email{venemans@mpia.de}
\altaffiltext{2}{Department of Physics, Winona State University, Winona, MN 55987, USA}
\altaffiltext{3}{Max-Planck-Institut f{\"u}r Radioastronomie, Auf dem H{\"u}gel 69, D-53121 Bonn, Germany}
\altaffiltext{4}{Department of Physics \& Astronomy, University of Wyoming, Laramie, WY 82070, USA}
\altaffiltext{5}{Institute of Astronomy, University of Cambridge, Madingley Road,
  Cambridge CB3 0HA, U.K.}
\altaffiltext{6}{Kavli Institute for Cosmology, University of Cambridge, Madingley Road,
  Cambridge CB3 0HA, U.K.}
\altaffiltext{7}{Astronomy Unit, Queen Mary University of London, London, E1 4NS, U.K.}
\altaffiltext{8}{Leiden Observatory, Leiden University, P.O. box 9513, NL-2300 RA Leiden, the Netherlands}

\begin{abstract}
We present ALMA band 3 observations of the CO(6--5), CO(7--6), and \ci\,369\,$\mu$m emission lines in three of the highest redshift quasar host galaxies at $6.6<z<6.9$. These measurements constitute the highest-redshift CO detections to date. The target quasars have previously been detected in \cii\,158\,$\mu$m emission and the underlying FIR dust continuum. We detect (spatially unresolved, at a resolution of $>$2\arcsec, or $\gtrsim$14\,kpc) CO emission in all three quasar hosts. In two sources, we detect the continuum emission around 400\,$\mu$m (rest-frame), and in one source we detect \ci\ at low significance. We derive molecular gas reservoirs of (1--3)$\times$10$^{10}$\,\msun\ in the quasar hosts, i.e.\ approximately only 10 times the mass of their central supermassive black holes. The extrapolated \cii\--to--CO(1--0) luminosity ratio is 2500--4200, consistent with measurements in galaxies at lower redshift. The  detection of the \ci\ line in one quasar host galaxy and the limit on the \ci\ emission in the other two hosts enables a first characterization of the physical properties of the interstellar medium in $z\sim7$ quasar hosts. In the sources, the derived global CO/\cii/\ci\ line ratios are consistent with expectations from photodissociation regions (PDR), but not X--ray dominated regions (XDR). This suggest that quantities derived from the molecular gas and dust emission are related to ongoing star-formation activity in the quasar hosts, providing further evidence that the quasar hosts studied here harbor intense starbursts in addition to their active nucleus. 
\end{abstract}


\section{Introduction}

Quasars are the most luminous, non-transient objects in the universe, and can be observed out to very high redshift \citep[e.g.,][]{fan06b,mor11,ven13,ban16}. In the case of quasars at $z>6$ (age of the Universe: $<$1\,Gyr), optical and near-infrared (NIR) observations trace the (rest-frame) UV emission of the quasars. These observations reveal that the quasars host supermassive black holes with masses exceeding $10^9$\,\msun\ in many cases \citep[e.g.,][]{jia07,kur07,der11,der14,wu15,ven15a}. Likewise, observations in the (sub-)mm regime have the potential to trace the rest-frame far-infrared (FIR) emission in the quasar host, as well as key diagnostic lines of their interstellar medium \citep[ISM, see, e.g.,][]{car13}.

Initial studies of the galaxies that host luminous quasars at $z\sim6$ unveiled that roughly 30\% of  quasars hosts were bright at mm wavelengths ($S_\mathrm{250\,GHz}\gtrsim1$\,mJy), implying FIR luminosities \lfir\,$\gtrsim3\times10^{12}$\,\lsun\ and star-formation rates (SFR) possibly exceeding 1000\,\msunyr \ \citep[e.g.,][]{ber03a,wan07,wan08b,wan16}. Follow-up studies of these FIR-bright quasar hosts targeting the redshifted CO emission line revealed that these galaxies have large reservoirs of cold molecular gas \citep[e.g.,][]{wal03,ber03b,car07,wan10,wan11a,wan11b,wan16}. These early studies of the host galaxies of $z\sim6$ quasars by nature concentrated on the FIR-bright quasars, and the results from these studies may introduce a biased view on the characteristics of the typical galaxy hosting a  $z\sim6$ quasar. This is supported by studies of lower luminosity $z\sim6$ quasars, powered by black holes with a mass of `only' $10^8$\,\msun, that reveal hosts with significantly fainter FIR luminosities of \lfir\,$\approx10^{11}$\,\lsun\ and SFRs of SFR\,$\ll100$\,\msunyr\ \citep{wil13,wil15}.

To extend the study of quasar host galaxies to $z\sim7$ we initiated a program targeting all quasars discovered at $z>6.5$ in \cii\ line emission and the underlying continuum, independent of their FIR brightness, with the aim of sampling the range of properties of quasar host galaxies and to investigate the star formation -- supermassive black hole growth relation at $z\sim7$.
The first quasar targeted in this project, J1120+0641 at $z=7.1$, already displayed a somewhat different characteristics in its FIR properties \citep[a fainter FIR luminosity of \lfir\,$=(6-18)\times10^{11}$\,\lsun\ and a more compact, $\sim$1\,kpc host galaxy;][]{ven12,ven17a} from the well-studied $z\sim6$ quasar hosts. Subsequent imaging of the FIR continuum and the \cii\,158\,$\mu$m emission line of additional $z>6.5$ quasars showed a range of properties with FIR luminosities from \lfir\,$\lesssim10^{12}$\,\lsun\ to \lfir\,$\sim10^{13}$\,\lsun\ \citep[][Mazzucchelli et al.\ in prep]{ban15b,ven16,ven17a,dec17a}. 

So far, our sample of $z>6.5$ quasar hosts have primarily been observed at 1\,mm targeting the \cii\ line and the underlying FIR continuum at 160$\mu$m in the rest-frame. To further study the characteristics of the interstellar medium (ISM) in $z>6.5$ quasar host galaxies, it is imperative to detect the FIR continuum of these galaxies at different frequencies and to observe additional molecular or atomic lines. In this paper, we present ALMA Cycle 2 observations of the CO(6--5), CO(7--6), and \ci(2--1) 369\,$\mu$m (hereafter \ci) emission line and the underlying dust continuum in three quasar host galaxies at $6.6<z<6.9$.
These quasar host galaxies are VIKING J030516.92--315056.0
(hereafter J0305--3150) at $z=6.6$, VIKING
J010953.13--304726.3 (hereafter J0109--3040) at $z=6.8$, and VIKING J234833.34--305410.0 (hereafter J2348--3054) at $z=6.9$, discovered in \citet{ven13}. At the time of discovery these three sources were the only quasars known at $z>6.5$ besides J1120+0641 at $z=7.1$ \citep{mor11}. The absolute magnitudes at 1450\,\AA\ in the rest-frame of these VIKING quasars are between --26.0 and --25.5, which is 1.4--1.9\,mag fainter than PSO~J036.5078+03.0498 at $z=6.54$, the most luminous $z>6.5$ quasar currently known \citep{ven15a} and $>$0.6\,mag brighter than the faint quasar J1205--0000 at $z\approx6.7$ discovered with the Subaru Hyper Suprime-Cam \citep{mat16}. The VIKING quasars are powered by black holes with masses of $\sim$$(1-2)\times10^9$\,\msun\ \citep{ven13,der14}. These black hole masses are comparable to those of other $z>6.5$ quasars \citep[e.g.,][Mazzucchelli et al.\ in prep]{mor11,der14,ven15a}. The three VIKING quasars were
previously observed with ALMA in Cycle 1 at $\sim$1\,mm targeting the redshifted \cii\ emission line \citep[presented in][]{ven16}. All three quasar host galaxies were detected  with \cii\
luminosities ranging between $(1.9-3.9)\times10^9$\,\lsun\ and
continuum luminosities at a rest-frame wavelength of 158\,$\mu$m of $\nu L_\nu
(158\,\mu$m$)=(0.8-4.3)\times10^{45}$\,erg\,s$^{-1}$ (see Table~\ref{tab:obsdesc}). The additional CO and \ci\ observations presented here, as well as the measurement of the underlying dust continuum emission at observed wavelengths around 3\,mm (rest-frame wavelength around 400\,$\mu$m), allow us, for the first time, to constrain the physical properties of quasar host galaxies at $z\sim7$ in more detail. 

The paper is organized as follows. In
Section~\ref{sec:observations} a description of the ALMA Cycle 2
observations is given. In Section~\ref{sec:luminosities} we present our results: in Sections~\ref{sec:j0305res}--\ref{sec:j2348res} we provide the derived properties for each of the three sources. In Section~\ref{sec:discussion} we discuss our results: in Section~\ref{sec:cont} we compare the dust continuum measurements made at 3\,mm (observed) with those at 1\,mm to constrain the shape of the dust spectral energy distribution and in Sections~\ref{sec:comass} and \ref{sec:cimass} we estimate the molecular gas and atomic carbon mass from the detected emission lines. In Section~\ref{sec:ism} we provide constraints on the properties of the interstellar medium (ISM), followed by a summary in Section~\ref{sec:summary}.

Throughout this paper, we adopt a concordance cosmology with parameters: $H_0=70$ km\,s$^{-1}$\,Mpc$^{-1}$, $\Omega_M=0.3$, and $\Omega_\Lambda=0.7$.

\section{ALMA OBSERVATIONS}
\label{sec:observations}

The three quasar host galaxies were observed in ALMA band-3 with 39--40 antennas in a compact configuration (baselines between 15 and 349\,m) between 2014 December 29 and 2015 January 6. A summary of the observations is given in Table~\ref{tab:obsdesc}. On 2014 December 29 J0109--3047 was observed for 47\,min (24\,min on-source), on 2014 December 30 observations of J2348--3054 were carried out for 60\,min of which 34\,min on-source, and on 2015 January 6 the host galaxy of J0305--3150 was observed for 31\,min (16\,min on-source). Our setup consisted of two pairs of two spectral windows, with each spectral window covering a frequency range of 1.875\,GHz at a resolution of 3.9\,MHz (11--13\,\kms). The two pairs of spectral windows are placed in sidebands that are separated by $\sim$12\,GHz. By fortuitous coincidence the frequency range that can be covered in this setup allows us to image two CO lines simultaneously for sources at $z\gtrsim6.5$. 

For all three quasar hosts the setup was tuned to include the CO(6--5) line in one of the four sidebands and the CO(7--6) and \ci\ lines in another sideband, using the redshift from the previous \cii\ observations. The two remaining spectral windows were placed between the CO(6--5) and CO(7--6) lines and utilized to increase the signal-to-noise of the continuum measurement. The beam size of $>$2\arcsec\ ($>$12\,kpc) ensured that the emission was likely unresolved by the ALMA observations as the maximum extent of the \cii\ emission in these sources is $<$0\farcs6 \citep{ven16}. For bandpass calibration, the sources J2258--2758, J2357--5311, and J0519--4546 were observed, respectively. The amplitude and flux calibration was performed through observations of the source J0334--401 and Mars, and the calibrators J2359--3133, J0120--2701, and J0334--4008, respectively, were observed every $\sim$7\,minutes for phase calibration. The raw data were reduced following standard reduction steps in the Common Astronomy Software Applications package \citep[CASA;][]{mul07}. The reduced cubes were cleaned with a weighting factor of robust = 2 (equivalent to natural weighting) to obtain the lowest noise maps. The rms noise per 100\,MHz bins averaged between 0.10\,mJy and 0.24\,mJy (Table~\ref{tab:obsdesc}). 

\begin{table*}[t]
\centering
\caption{Properties of the Observed Quasar Host Galaxies, Description of the ALMA Cycle 2 Observations, and the Derived Characteristics of the Hosts \label{tab:obsdesc}}
\begin{tabular}{lccc}
\hline 
\hline
{} & J0305--3150 & J0109--3047 & J2348--3054 \\
\hline
R.A.\ (J2000) & 03$^\mathrm{h}$05$^\mathrm{m}$16$^\mathrm{s}\!\!$.91 
& 01$^\mathrm{h}$09$^\mathrm{m}$53$^\mathrm{s}\!\!$.13 
& 23$^\mathrm{h}$48$^\mathrm{m}$33$^\mathrm{s}\!\!$.35 \\
Decl.\ (J2000) & --31$^\circ$50\arcmin55\farcs94 
& --30$^\circ$47\arcmin26\farcs32 
& --30$^\circ$54\arcmin10\farcs30 \\
$z_\mathrm{[CII]}$ & 6.6145$\pm$0.0001 
& 6.7909$\pm$0.0004 
& 6.9018$\pm$0.0007 \\
\lcii\ (\lsun) & $(3.9\pm0.2)\times10^9~$ 
& $(2.4\pm0.2)\times10^9~$ 
& $(1.9\pm0.3)\times10^9~$ \\
FWHM$_\mathrm{[CII]}$ (\kms) & 255$\pm$12~
& 340$\pm$34~ 
& 405$\pm$69~  \\
\lfir\ (\lsun)\tablenotemark{a} & $(7.3_{-3.3}^{+0.2})\times10^{12}$  
& $(1.3_{-0.7}^{+0.2})\times10^{12}$ 
& $(4.5_{-2.3}^{+0.4})\times10^{12}$ \\
$M_d$ (\msun)\tablenotemark{a} & $(4.5-24)\times10^8$
& $(0.7-4.9)\times10^8$
& $(2.7-15)\times10^8$ \\
\hline
$\nu_\mathrm{obs}$ (GHz) & 90.6--94.4, 102.6--106.5 
& 88.4--92.2, 100.4--104.2 
& 87.0--90.8, 99.0--102.9 \\
$t_\mathrm{exp,on-source}$ (min) & 16 & 24 & 34 \\
\# of antennas & 39 & 40 & 39 \\
RMS noise (per 100\,MHz) & 242\,$\mu$Jy & 121\,$\mu$Jy & 102\,$\mu$Jy \\
beam size & 5\farcs3\,$\times$\,2\farcs2 
& 4\farcs2\,$\times$\,2\farcs5
& 3\farcs8\,$\times$\,2\farcs3 \\ 
\hline
CO(6--5) flux (\jykms) & $0.65\pm0.07$ & $0.11\pm0.05$ & $0.28\pm0.05$ \\
$L_\mathrm{CO(6-5)}$ (\lsun) & $(2.6\pm0.3)\times10^{8}$~ 
& $(4.4\pm1.9)\times10^{7}$~ 
& $(1.2\pm0.2)\times10^{8}$~ \\
$L^\prime_\mathrm{CO(6-5)}$ (\kkmspc) & $(2.6\pm0.3)\times10^{10}$ 
& $(4.5\pm2.0)\times10^9$~ 
& $(1.2\pm0.2)\times10^{10}$ \\
CO(7--6) flux (\jykms) & $0.69\pm0.09$ & $0.24\pm0.04$ & $0.26\pm0.06$ \\
$L_\mathrm{CO(7-6)}$ (\lsun) & $(3.2\pm0.4)\times10^{8}$~ 
& $(1.2\pm0.2)\times10^8$~ 
& $(1.3\pm0.3)\times10^8$~ \\
$L^\prime_\mathrm{CO(7-6)}$ (\kkmspc) & $(2.0\pm0.3)\times10^{10}$ 
& $(7.5\pm1.3)\times10^9$~ 
& $(8.1\pm1.7)\times10^9$~ \\
\ci(2--1) flux (\jykms) & $<0.35$ & $<0.15$ & $0.16\pm0.06$ \\
$L_\mathrm{[CI](2-1)}$ (\lsun) & $<1.6\times10^{8}$~ & $<7.4\times10^7$~ & 
$(8.0\pm2.8)\times10^7$~  \\
$L^\prime_\mathrm{[CI](2-1)}$ (\kkmspc) & $<9.6\times10^{9}$~ & $<4.4\times10^9$~ & 
$(4.7\pm1.7)\times10^9$~  \\
$S_\mathrm{3\,mm}$ ($\mu$Jy) & $233\pm30$~ & $<46$ & $118\pm13$~ \\
\hline
$L^\prime_\mathrm{CO(1-0)}$\tablenotemark{b} (K\,\kms\,pc$^2$) & $(3.4\pm0.3)\times10^{10}$ & 
$(1.3\pm0.2)\times10^{10}$ & $(1.4\pm0.2)\times10^{10}$ \\
\lcii/$L_\mathrm{CO(1-0)}$\tablenotemark{b} 
& {2530$\pm$130~} & {4170$\pm$350~} & {2860$\pm$450~} \\
$M_{\mathrm{H}_2,\mathrm{CO}}$ (\msun)\tablenotemark{c} & $(2.7\pm0.2)\times10^{10}$ & $(1.0\pm0.2)\times10^{10}$ & 
$(1.2\pm0.2)\times10^{10}$ \\
$M_{\mathrm{H}_2,\mathrm{dust}}$ (\msun)\tablenotemark{d} & $(2.4-18)\times10^{10}$ & $(0.4-3.7)\times10^{10}$ 
& $(1.4-11)\times10^{10}$ \\
$M_\mathrm{CI}$ (\msun)\tablenotemark{e} & $<2.1\times10^7$ & $<9.6\times10^6$ & $(1.0\pm0.4)\times10^{7~}$ \\
\hline
\end{tabular}
\tablenotetext{a}{\lfir\ and $M_d$ derived from the continuum detection at 1\,mm, taken from \citet{ven16}.}
\tablenotetext{b}{A CO excitation ladder similar to that of the $z=6.4$ quasar J1148+5251 is assumed (see Section~\ref{sec:comass}).}
\tablenotetext{c}{Molecular gas mass derived from the CO(1--0) luminosity, assuming a luminosity-to-gas mass conversion factor of $\alpha=0.8$\,\msun\,(\kkmspc)$^{-1}$.}
\tablenotetext{d}{Molecular gas mass derived from the dust mass, assuming a gas-to-dust mass ratio of 70--100 and a molecular gas mass fraction of 0.75 (see Section\,\ref{sec:comass}).}
\tablenotetext{e}{Assuming an excitation temperature of $T_\mathrm{ex}=30$\,K (see Section~\ref{sec:cimass}).}
\end{table*}

\begin{figure*}
\includegraphics[width=\textwidth]{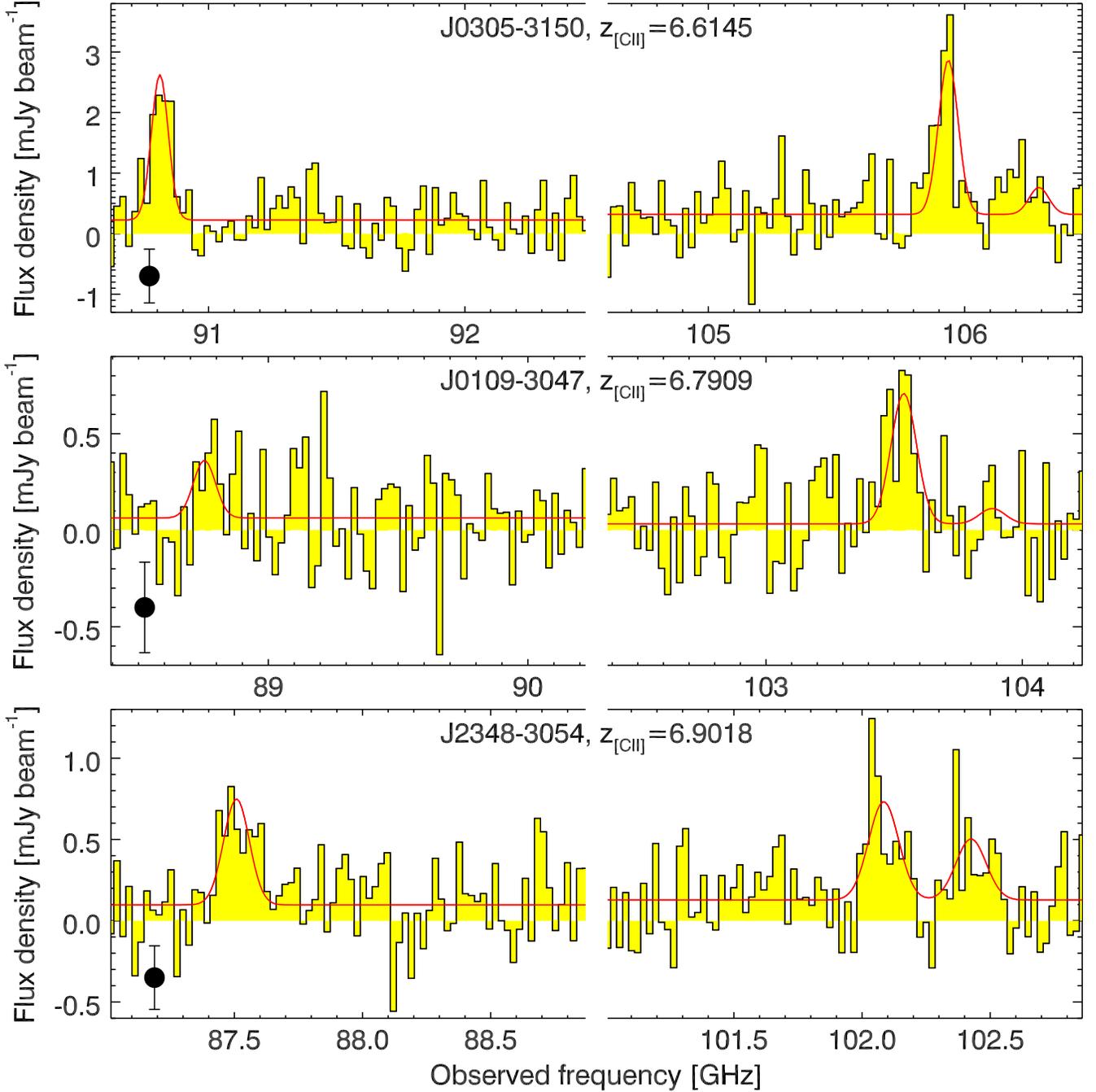}
\caption{ALMA 3\,mm spectra containing the CO and \ci\ emission lines of the three $6.6<z<6.9$ quasar host galaxies, extracted at the position of the \cii\ emission (that is coincident with the quasar position). The channels were binned by a factor of 6 to a width of 23.4\,MHz ($\sim$80\,\kms). The typical uncertainty per bin is shown in the lower left corner. The red solid line shows a fit to the CO(6--5), CO(7--6), and \ci\ lines with the redshift and line width fixed to those of the \cii\ emission line (see Table~\ref{tab:obsdesc}). }
\label{fig:spectra}
\end{figure*}

\begin{figure*}
\includegraphics[width=\textwidth]{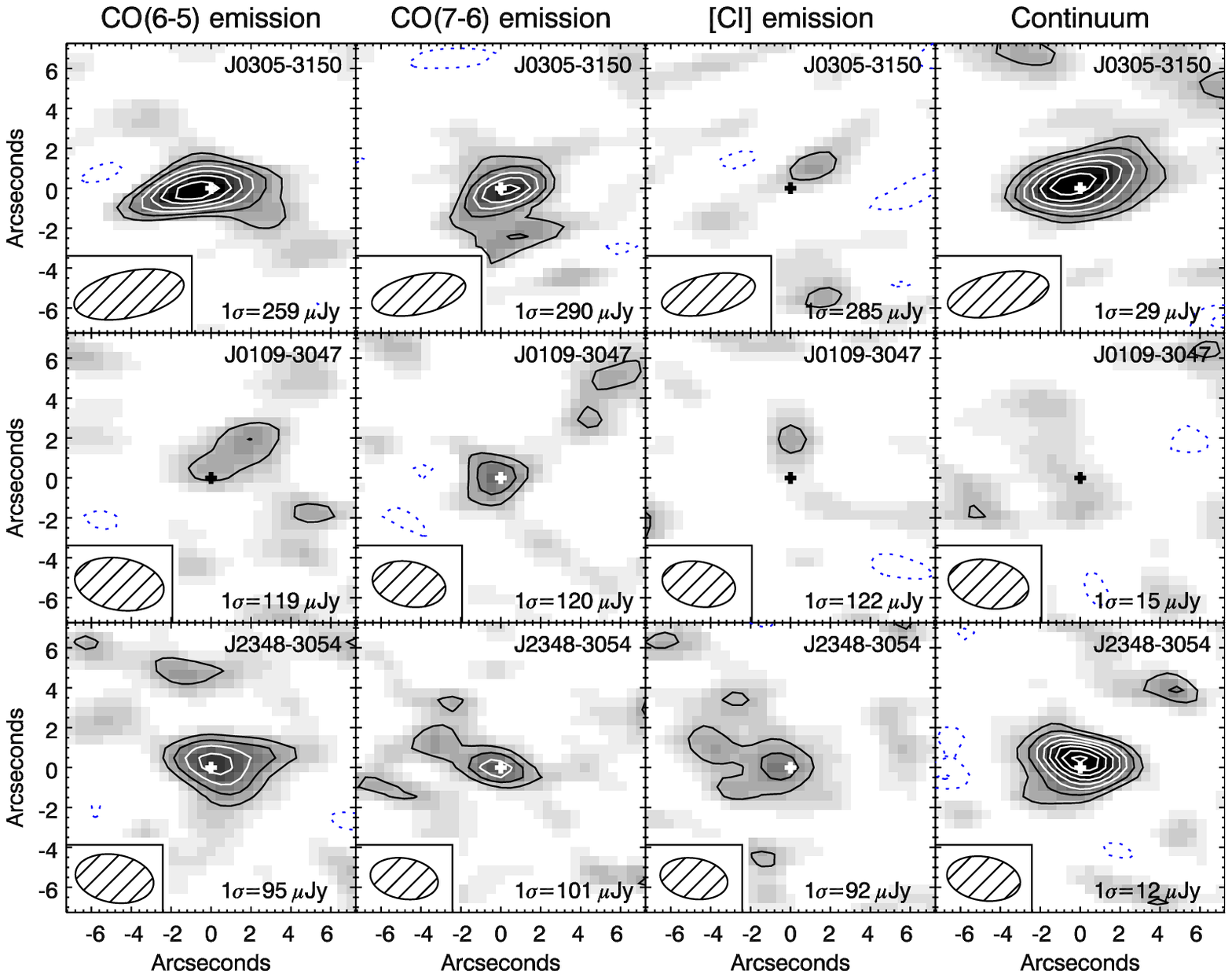}
\caption{Maps of the line and continuum emission from the three quasar host galaxies. From left to right the CO(6--5), CO(7--6), \ci, and underlying continuum emission are shown. To create the line emission maps, the data cubes were first continuum subtracted and subsequently averaged over the FWHM of the \cii\ line (405\,\kms, 340\,\kms, and 255\,\kms\ for J2348--3054, J0109--3047, and J0305--3150, respectively, see Table~\ref{tab:obsdesc}). The beam is shown in the bottom left of each map and in the bottom right the 1\,$\sigma$ rms noise is printed. The blue, dashed contours are --3$\sigma$ and --2$\sigma$, the black, solid contours are +2$\sigma$ and +3$\sigma$, and the white solid contours are [4, 5, 6, 7, 8, 9] $\times \sigma$. The small white and black crosses indicate the position of the \cii\ emission of quasar host galaxies.}
\label{fig:maps}
\end{figure*}

\section{CO and \ci\ Spectra}
\label{sec:luminosities}

In Figure~\ref{fig:spectra} we show the CO and \ci\ spectra of the quasar hosts extracted at
the position of the center of the \cii\ emission. We fitted a Gaussian function
to the lines, fixing the redshift and the width of the lines to those from the
significantly higher signal-to-noise (S/N) \cii\ line \citep{ven16}. In Figure~\ref{fig:maps} we averaged the data at the redshift
given by the \cii\ line over the width of the \cii\ line. Channels not
belonging to emission lines were averaged to create continuum maps. 
Below we will describe the results for each of the sources individually.

\subsection{J0305--3150 ($z_\mathrm{[CII]}=6.6145$)}
\label{sec:j0305res}

The host galaxy of quasar J0305--3150 shows the brightest \cii\ line of
the three VIKING quasars presented in \citet{ven16}. The \cii\ line was
detected at a redshift of $z_\mathrm{[CII]}=6.6145\pm0.0001$ with a strength
of $F_\mathrm{[CII]}=3.44\pm0.15$\,\jykms\ and a width of
FWHM\,=\,$255\pm12$\,\kms. Fitting a Gaussian with a FWHM of 255\,\kms\
centered on the (redshifted) frequency of the CO(6--5) emission line
($\nu_\mathrm{CO(6-5),obs}=\nu_\mathrm{CO(6-5),rest}/(1+z_\mathrm{[CII]})$) to
the spectrum (Figure~\ref{fig:spectra}) resulted in $\sim$8$\sigma$ detection
(Figure~\ref{fig:maps}) of the CO(6--5) line with a strength of
$0.65\pm0.07$\,\jykms\ and a luminosity of \lcovi\,$=(2.6\pm0.3)\times10^8$\,\lsun. 
Allowing the width and center of the
Gaussian to vary gives very similar parameters:
$z_\mathrm{CO(6-5)}=6.6139\pm0.0005$, FWHM$_\mathrm{CO(6-5)}=314\pm48$\,\kms,
and $F_\mathrm{CO(6-5)}=0.74\pm0.10$\,\jykms. 

The CO(7--6) emission line was detected at $\sim$6$\sigma$ with
$z_\mathrm{CO(7-6)}=6.6155\pm0.0004$, FWHM$_\mathrm{CO(7-6)}=225\pm38$\,\kms,
and $F_\mathrm{CO(7-6)}=0.68\pm0.10$\,\jykms. Forcing the line to have the
same width and redshift as the \cii\ line results in a similar line strength as above
of $0.69\pm0.09$\,\jykms\ and a CO(7--6) luminosity of \lcovii\,$=(3.2\pm0.4)\times10^8$\,\lsun. 
The width, redshift, and spatial
location of the CO(6--5) and CO(7--6) emission are, within the uncertainties,
consistent with those of the \cii\ line. This indicates that gas components
traced by the \cii\ and CO emission have, on average, similar kinematics. 

The measured CO fluxes of $F_\mathrm{CO(6-5)}=0.65$\,\jykms\ and
$F_\mathrm{CO(7-6)}=0.69$\,\jykms\ make the host of J0305--3150 roughly as bright as 
the archetypical, luminous SDSS quasar J1148+5251 (which has $F_\mathrm{CO(6-5)}=0.67$\,\jykms\ and $F_\mathrm{CO(7-6)}=0.63$\,\jykms, \citealt{wal03,ber03b,rie09}).

The \ci\ emission line has not been significantly detected in both the spectrum and the line map
(Figures~\ref{fig:spectra} and \ref{fig:maps}). At the position and redshift of
the \cii\ emission we measure a 3$\sigma$ upper limit to the line flux of
$F_\mathrm{[CI]}<0.35$\,\jykms\ ($L_\mathrm{[CI]}<1.6\times10^8$\,\lsun)

Averaging the frequency channels not covered by emission lines resulted in a
significant, 8$\sigma$ detection of the continuum at an observed frequency of 98.4\,GHz (rest-frame frequency of
749\,GHz, rest-frame wavelength of $\sim$400\,$\mu$m) of
$S_\mathrm{98.4\,GHz}=233\pm30$\,$\mu$Jy. We will
discuss the implications of this detection further in Section~\ref{sec:cont}.

\subsection{J0109--3047 ($z_\mathrm{[CII]}=6.7909$)}
\label{sec:j0109res}

The host galaxy of quasar J0109--3047 was detected in \cii\ with
$z_\mathrm{[CII]}=6.7909\pm0.0004$, $F_\mathrm{[CII]}=2.04\pm0.20$\,Jy\,\kms,
and a width of FWHM\,$=340\pm36$
\citep{ven16}. The continuum of the host was the faintest of the three quasars
considered here with $S_\mathrm{1\,mm}=0.56\pm0.11$\,mJy at 158\,$\mu$m in the rest-frame.

The CO(6--5) line is only marginally detected in this quasar host with a
S/N$\approx$2.3 (Figures~\ref{fig:spectra} and \ref{fig:maps}). From a fit to the spectrum, we derive a line flux of $F_\mathrm{CO(6-5)}=0.11\pm0.05$\,\jykms\ and a luminosity of \lcovi\,$=(4.4\pm1.9)\times10^7$\,\lsun. In contrast, the CO(7--6) line was
detected with a S/N\,$\approx6$ at the same spatial position as the \cii\ line in
the line map (Figure~\ref{fig:maps}). Fitting a Gaussian to the spectrum while
fixing the width and redshift to that of the \cii\ line gives a line strength of
$F_\mathrm{CO(7-6)}=0.24\pm0.04$\,\jykms (\lcovii\,$=(1.2\pm0.2)\times10^8$\,\lsun). We do not detect the \ci\ line and
place 3$\sigma$ limits on the line flux of $F_\mathrm{[CI]}<0.15$\,\jykms\ ($L_\mathrm{[CI]}<7.4\times10^7$\,\lsun).

The host galaxy of J0109--3047 was not detected in the 3\,mm continuum image
(Figure~\ref{fig:maps}). At the position of the quasar host galaxy we measure a
continuum flux density of $S_\mathrm{96.3\,GHz}=22\pm15$\,$\mu$Jy (corresponding to a rest-frame frequency of 750\,GHz). The 3$\sigma$
upper limit for the continuum flux density is 46\,$\mu$Jy.

\subsection{J2348--3054 ($z_\mathrm{[CII]}=6.9018$)}
\label{sec:j2348res}

J2348--3054 is the highest redshift quasar in our sample with
$z_\mathrm{[CII]}=6.9018\pm0.0007$. It has the faintest \cii\ line of our
three observed quasars with $F_\mathrm{[CII]}=1.57\pm0.26$\,Jy\,\kms,
and FWHM$_\mathrm{[CII]}=405\pm69$\,\kms\ \citep{ven16}. 

Both the CO(6--5) and CO(7--6) lines are detected at S/N$\sim$5 in the ALMA
band 3 data (Figures~\ref{fig:spectra} and \ref{fig:maps}). The line fluxes
measured from the spectrum, using the \cii\ redshift and line width, are
$F_\mathrm{CO(6-5)}=0.28\pm0.05$\,\jykms\ and
$F_\mathrm{CO(7-6)}=0.26\pm0.06$\,\jykms. This corresponds to line luminosities of \lcovi\,$=(1.2\pm0.2)\times10^8$\,\lsun\ and \lcovii\,$=(1.3\pm0.3)\times10^8$\,\lsun which is very similar to the
CO(7--6) luminosity of J0109--3047.

Intriguingly, the \ci\ line in J2348--3054 was also detected albeit with a low significance
(S/N$\sim$3). Both in frequency (Figure~\ref{fig:spectra}) and
spatially (Figure~\ref{fig:maps}) the \ci\ emission coincides with the expectations from the \cii\
emission line. We derive a line flux of $F_\mathrm{[CI]}=0.16\pm0.06$\,\jykms\ and a luminosity of $L_\mathrm{[CI]}=(8.0\pm2.8)\times10^7$\,\lsun\ from the spectrum. We
will discuss the implications of the detection of this line in
Section~\ref{sec:cimass}.

The continuum of J2348--3054 was also detected (Figure~\ref{fig:maps}) with
$S_\mathrm{94.9\,GHz}=118\pm13$\,$\mu$Jy (S/N$\sim$9).

\section{DISCUSSION}
\label{sec:discussion}

\subsection{Constraints on the Dust Emission}
\label{sec:cont}

\begin{figure*}
\plottwo{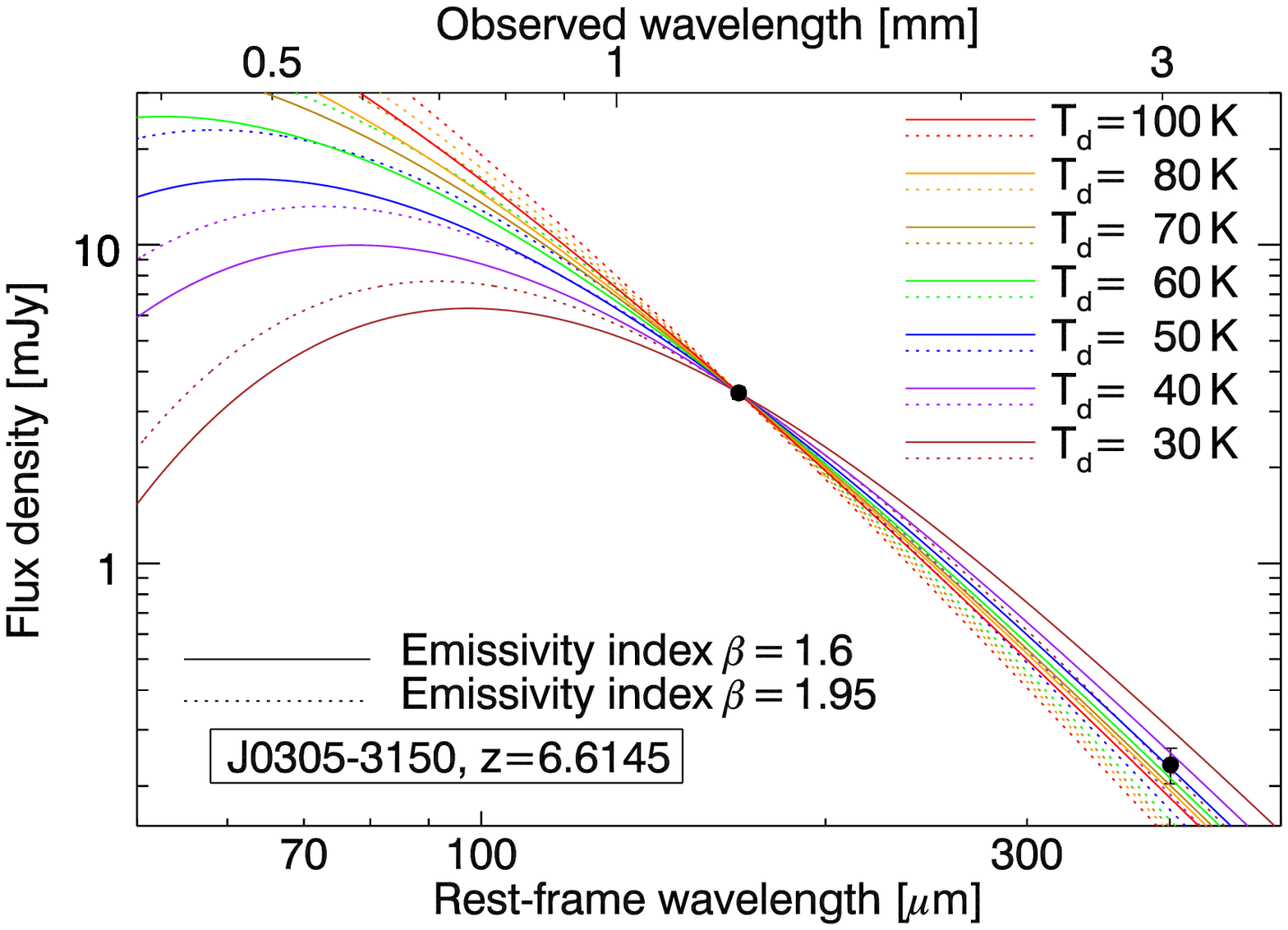}{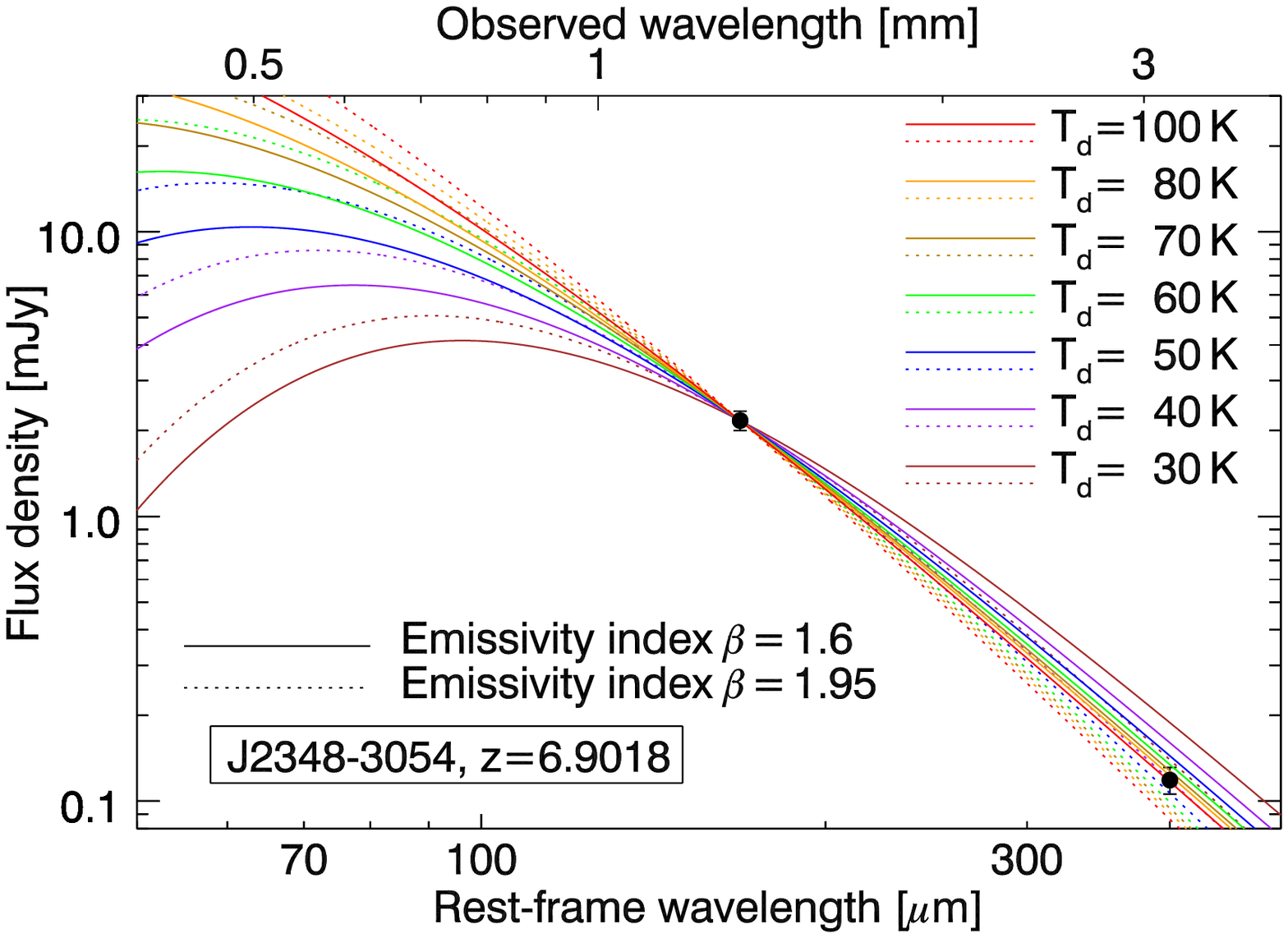}
\caption{Measured flux densities at observed wavelengths around 1\,mm and 3\,mm of the far-infrared dust continuum in J0305--3150 ({\em left}) and J2348--3054 ({\em right}). Overplotted are modified black body curves with different temperatures for two dust emissivity indices ($\beta=1.6$ and $\beta=1.95$) fixed to the data point at 1.3\,mm. The solid lines show the observed dust SED assuming $\beta=1.6$, while the dashed line assumes an emissivity index of $\beta=1.95$. With $\beta$ fixed to $\beta=1.6$ we derive best-fitting temperatures of $T_d=47^{+21}_{-10}$\,K for J0305--3150 and $T_d=94^{+174}_{-35}$\,K for J2348--3054. For $\beta=1.95$, we find $T_d=28^{+7}_{-5}$\,K and $T_d=40^{+13}_{-8}$\,K for J0305--3150 and J2348--3054, respectively. The upper limit on the 3\,mm continuum strength that we derive for J0109--3047 results in a lower limit of $T_d>27$\,K in this quasar host (Section~\ref{sec:cont}). To better constrain the temperature, continuum detections at smaller wavelengths are required.}
\label{fig:contfit}
\end{figure*}

Analysis of the far-infrared emission in luminous, high redshift ($z\gtrsim2$) quasars revealed that the dust in these objects has typical temperatures between 40\,K and 60\,K with a mean of $T_d=47\pm3$\,K \citep[e.g.,][]{pri01,bee06,lei14}. A dust temperature of $T_d=47$\,K has subsequently often been assumed for studies of the cold dust emission of $z\gtrsim6$ quasar hosts \citep[e.g.,][]{wil13,wil15,wan13,ban15b,ven16}. In this section we combine the continuum measurements from our ALMA 1\,mm
data (presented in \citealt{ven16}) with the continuum detections in the 3\,mm data presented here to test whether the dust temperature is consistent with the canonical value. The addition of a
continuum point at 3\,mm significantly increases the baseline over which we can
constrain the dust SED. However, both our continuum points are on the Rayleigh-Jeans tail of the dust emission and we cannot tightly constrain the dust temperature at these relatively long wavelengths. Furthermore, we need to make assumptions about the properties of the dust. Following the literature, we here assume that the dust emission can be described by a modified black body with a dust temperature $T_d$ and a power-law emissivity index $\beta$ \citep[e.g.,][]{pri01,ven16}. With only two continuum detections (at $\sim$1\,mm and $\sim$3\,mm), we cannot constrain $T_d$ and $\beta$ at the same time. In the following discussion, we will assume two different values of $\beta$ from the literature: $\beta=1.6$ \citep{bee06} and $\beta=1.95$ \citep{pri01}.

As discussed in \citet{ven16}, it is important to take
the effects of the CMB into account, which has a temperature of
$T_\mathrm{CMB}\approx21$\,K at $z=6.8$. The CMB provides both an additional
source of heating and a background that reduces the detectability of the
emission from the quasar hosts, see \citet{dac13} for an extensive discussion
on these effects. While the heating by the CMB is negligible for our sources
if the dust temperature is $T_d\gtrsim30$\,K, the CMB can significantly reduce the flux density we measure
from a source at high redshift, especially at the low frequencies:

\begin{equation}
S_\nu^{\mathrm{obs}}/S_\nu^{\mathrm{intrinsic}} =
1-B_\nu[T_\mathrm{CMB}(z)]/B_\nu[T_d],
\label{eq:cmbc}
\end{equation}

\noindent
with $B_\nu$ the Planck function at
rest-frame frequency $\nu$ \citep{dac13}. I.e., with a dust temperature of $T_d=30$\,K
and a redshift of $z=6.6$, we are only measuring 75\% and 50\% of the
intrinsic flux density at rest-frame wavelengths of 158\,$\mu$m and
400\,$\mu$m, respectively.

To compare the continuum flux densities at 1\,mm with those measured at 3\,mm, we first averaged the line-free sidebands at 1\,mm and subsequently convolved the resulting image to the same resolution as the 3\,mm data. 
For J0305--3150 we measure $S_\mathrm{234.6\,GHz}=3.43\pm0.37$\,mJy.
The observed flux ratio in the two ALMA bands is thus $S_\mathrm{1\,mm}/S_\mathrm{3\,mm}=14.7$. This is very similar to the ratio of $S_\mathrm{1\,mm}/S_\mathrm{3\,mm}\sim14$ measured in the well-studied quasar host J1148+5251 at $z=6.42$ \citep{rie09,gal14}. For J2348--3054 we measure $S_\mathrm{225.4\,GHz}=2.17\pm0.17$\,mJy and $S_\mathrm{1\,mm}/S_\mathrm{3\,mm}=18.3$.

In Figure~\ref{fig:contfit} we show the continuum detections of the host
galaxies of J0305--3150 and J2348--3054. We
fitted a modified black body to the data points, taking into account the effects of the
CMB described above, with a fixed redshift of $z=6.6145$ and $z=6.9018$ respectively and two different dust emissivity indices of $\beta=1.6$ and $\beta=1.95$. We added an uncertainty of 10\% in quadrature to account for the absolute flux calibration uncertainty. We derive a dust temperature of $T_d=47^{+21}_{-10}$\,K ($T_d=28^{+7}_{-5}$\,K) assuming a $\beta=1.6$ ($\beta=1.95$) for J0305--3150 and $T_d=94^{+174}_{-35}$\,K ($T_d=40^{+13}_{-8}$\,K) for J2348--3054. Within the large uncertainties, these values are consistent with the canonical values of $T_d=41-47$\,K. If instead we fix the dust temperature to $T_d=47$\,K, we derive values for the dust emissivity index of $\beta=1.60^{+0.16}_{-0.15}$ for J0305--3150 and $\beta = 1.86\pm0.15$ for J2348--3047. As is demonstrated in Figure~\ref{fig:contfit}, our observed continuum flux densities at 1\,mm and 3\,mm do not pose tight constraints on the dust temperature and emissivity index. To better constrain the dust temperature, we need to measure the dust continuum at rest-frame wavelengths $<$100\,$\mu$m in these quasar hosts. This is feasible with, for example, ALMA band 8 and 9 observations.

The 3\,mm continuum was not detected in the host galaxy of quasar J0109--3047,
and we therefore could not constrain the dust temperature in this source. By
taking the 3$\sigma$ upper limit of $S_{3\,\mathrm{mm}}=0.046$\,mJy, we
derived a lower limit on the dust temperature of $T_d>27$\,K. For a dust temperature of $T_d=47$\,K and $\beta=1.6$, we expect, based on the continuum detection at $\sim$1\,mm, a 3\,mm continuum flux density of $S_\mathrm{96.3\,GHz}\approx33$\,$\mu$Jy, which is consistent with our observed limits. 

In the remainder of this paper, we will use the FIR luminosities derived from the continuum detection at 1\,mm by \citet{ven16}. These values are listed in Table~\ref{tab:obsdesc}. The FIR luminosity is obtained by assuming $T_d=47$\,K and $\beta=1.6$, while the error bar includes both the measurement error and the uncertain shape of FIR continuum. The latter is determined by scaling model templates of local star-forming galaxies to the continuum detection, see \citet{ven16} for details.

\subsection{Molecular Gas Mass Derived from CO}
\label{sec:comass}

The mass of the molecular (mostly H$_2$) gas can be estimated using the
equation $M_\mathrm{gas}=\alpha L^\prime_\mathrm{CO(1-0)}$ with
$M_\mathrm{gas}$ the molecular gas mass, $\alpha$ the CO luminosity-to-gas
mass conversion factor, and $L^\prime_\mathrm{CO(1-0)}$ the CO(1--0) luminosity
in units of K\,\kms\,pc$^2$. The luminosity of an emission line can be derived by

\begin{eqnarray}
L_\mathrm{line}^\prime &=& L_\mathrm{line}/(3\times10^{-11}\,\nu_\mathrm{rest}^3) = \nonumber \\
&=& 3.25\times10^7\,F_\mathrm{line}\,D_L^2\,(1+z)^{-3}\,\nu_\mathrm{obs}^{-2}, 
\label{eq:colum}
\end{eqnarray}

\noindent
with $\nu_\mathrm{rest}$ and $\nu_\mathrm{obs}$ the rest-frame and observed frequency of
the emission line in GHz, $D_L$ the luminosity distance in Mpc and $F_\mathrm{line}$ the line flux in Jy\,\kms\ \citep[e.g.,][]{car13}. Following the literature, we further assume
a value of $\alpha=0.8$\,\msun\,(K\,\kms\,pc$^2$)$^{-1}$, derived for local
ultra-luminous infrared galaxies \citep[ULIRGs; e.g.,][]{dow98}. This value is also used to compute molecular gas masses in $z\sim6$ quasar host galaxies \citep[e.g.,][]{wan10}.

Since we only measured the CO(6--5) and/or CO(7--6) line flux in our sources, we have to assume
a CO spectral line energy distribution (CO SLED) to estimate the CO(1--0) line strength. The CO emission of distant quasars peaks around that of CO(6--5) and CO(7--6) \citep[e.g.,][]{rie09,car13}. To estimate CO(1--0) luminosity in our sources, we here apply the same model that fits the strength of the CO lines of the quasar host galaxy J1148+5251 at $z=6.42$ \citep[][]{rie09,stef15}. In J1148+5251 several low-J and high-J CO lines have been detected, including the CO(2--1), CO(3--2), CO(6--5), and CO(7--6) lines \citep[][]{ber03b,wal03,rie09,stef15}. The large velocity gradient (LVG) model by \citet{rie09} that fits the observed CO line fluxes in J1148+5251 results in observed CO(6--5) and CO(7--6) line fluxes that are of 
roughly equal strength, $F_\mathrm{CO(7-6)}/F_\mathrm{CO(6-5)}=1.03$, and line flux ratios of $F_\mathrm{CO(6-5)}/F_\mathrm{CO(1-0)}=28$ and $F_\mathrm{CO(7-6)}/F_\mathrm{CO(1-0)}=29$. Following Equation~(\ref{eq:colum}), the CO luminosity ratios of the LVG model are $L^\prime_\mathrm{CO(6-5)}/L^\prime_\mathrm{CO(1-0)}\approx0.78$ and $L^\prime_\mathrm{CO(7-6)}/L^\prime_\mathrm{CO(1-0)}\approx0.60$. Note that in the two cases where we detect both CO(6--5) and CO(7--6) lines, the ratio is indeed close to 1 (Table~\ref{tab:obsdesc}): $F_\mathrm{CO(7-6)}/F_\mathrm{CO(6-5)}\approx1.06$ and $F_\mathrm{CO(7-6)}/F_\mathrm{CO(6-5)}\approx0.93$ for J0305--3150 and J2348--3054, respectively.

Using Equation~(\ref{eq:colum}), we derive CO(6--5) and CO(7--6) luminosities in our quasar hosts of \lpcovi\,$=(0.45-2.6)\times10^{10}$\,\kkmspc\ and \lpcovii\,$=(0.75-2.0)\times10^{10}$\,\kkmspc\ (see Table~\ref{tab:obsdesc}). Assuming the CO excitation ladder in the VIKING quasar hosts can be described with the one derived for J1148+5251, we can obtain $L^\prime_\mathrm{CO(1-0)}$ luminosities by applying the \lpcovi/$L^\prime_\mathrm{CO(1-0)}$ and \lpcovii/$L^\prime_\mathrm{CO(1-0)}$ luminosity ratios given by the LVG model. For J0305--3150 and J2348--3054, where we detect both CO(6--5) and CO(7--6) lines at S/N\,$>3$, we take the weighted mean of the two estimated CO(1--0) luminosities. For J0109--3047 we only consider the CO(7--6) line as the CO(6--5) line has a low significance (Figures~\ref{fig:spectra} and \ref{fig:maps}). The CO(1--0) luminosities we derive for the quasar hosts are $L^\prime_\mathrm{CO(1-0), J0305-3150}=(3.4\pm0.4)\times10^{10}$, $L^\prime_\mathrm{CO(1-0), J0109-3047}=(1.3\pm0.2)\times10^{10}$, and $L^\prime_\mathrm{CO(1-0), J2348-3054}=(1.4\pm0.2)\times10^{10}$\,\kkmspc.

\begin{figure}
\includegraphics[width=\columnwidth]{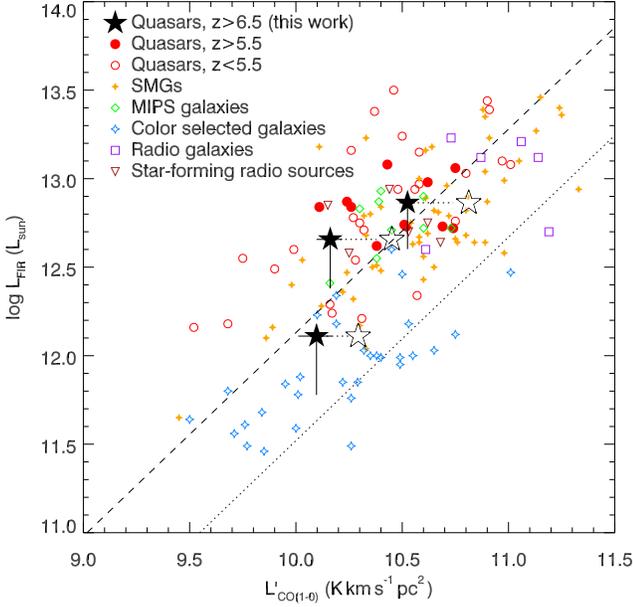}
\caption{FIR luminosity as function of $L^\prime_\mathrm{CO(1-0)}$ for various types of objects at $z>1$ (adapted from \citealt{car13}). The small, colored symbols are sources with a CO detection at $z>1$, compiled by \citet{car13}. Plotted are CO and FIR luminosities of star-forming radio sources (upside-down triangles), radio galaxies (squares), color selected galaxies at $z>1$ (small open stars), MIPS 24$\mu$m selected galaxies (diamonds), sub-millimeter galaxies (SMGs, small filled stars), quasars at $z<5.5$ (open circles), and quasars at $5.5<z<6.5$ (filled circles). Our detections of the $z>6.5$ quasar hosts are plotted as large, filled stars. The uncertainties in $L^\prime_\mathrm{CO(1-0)}$ only reflect the measurement errors and do not include the uncertainty in the CO SLED. This is illustrated by the large, open stars that show the derived $L^\prime_\mathrm{CO(1-0)}$ luminosities assuming a different CO excitation ladder (see Section~\ref{sec:comass} for details). The dashed (dotted) line indicates the relation between \lfir\ and  $L^\prime_\mathrm{CO(1-0)}$ derived for starburst (star-forming) galaxies up to $z\sim3.5$ from e.g.,\citet{dad10} and \citet{gen10}.}
\label{fig:lco}
\end{figure}

The \cii/CO(1--0) luminosity ratios derived for three quasar hosts range from 2500 to 4200. These values are within a factor of 2 of the ratio of \lcii/$L_\mathrm{CO(1-0)} \approx 4100$ measured in local starburst galaxies and star forming regions in the Milky Way \citep[e.g.,][]{sta91} and in dusty star-forming galaxies at $z>2$ \citep[e.g.,][]{gul15}. 
In Figure~\ref{fig:lco} we plot the FIR luminosity against CO(1--0) luminosity $L^\prime_\mathrm{CO(1-0)}$ of our quasar hosts. The CO and FIR luminosities of J0305--3150 and J2348--3054 are very similar to those in $z\sim6$ quasar host galaxies \citep[e.g.,][]{wan10,wan11a,wan11b}. The $z>6.6$ host galaxies discussed here have CO(1--0) over FIR luminosity ratios consistent with those of starburst galaxies at $z=0-3.5$ \citep[e.g.,][]{dad10,gen10}, see Figure~\ref{fig:lco}. 

Based on the derived CO(1--0) luminosities, we estimate that the quasar host galaxies contain a molecular gas mass of $(2.7\pm0.2)\times10^{10}$\,\msun\ (J0305--3150), $(1.0\pm0.2)\times10^{10}$\,\msun\ (J0109--3047), and $(1.2\pm0.2)\times10^{10}$\,\msun\ (J2348--3054, see Table~\ref{tab:obsdesc}). We can compare these gas masses with the dynamical mass derived from the \cii\ emission line for these quasar host galaxies in \citet{ven16}. For J0109--3047 and J2348--3054 the dynamical masses are $(1.4\pm0.4)\times10^{11}$\,\msun\ and $(7.2\pm3.6)\times10^{10}$\,\msun, and roughly 4\%--39\% of that dynamical mass is comprised of molecular gas. Assuming dark matter does not significantly contribute to the mass in the centre of these galaxies \citep[e.g.,][]{gen17}, this suggests that these host galaxies contain a large stellar mass of (at most) $M_*\approx M_\mathrm{dyn}-M_{\mathrm{H}_2}=(2-17)\times10^{10}$\,\msun, which is at the high end of the stellar mass function derived for star-forming galaxies at similar redshifts \citep[e.g.,][]{bow14,gra15,ste15}. On the other hand, in J0305--3150 as much as 54--81\% of the dynamical mass of $M_\mathrm{dyn}=(4.1\pm0.5)\times10^{10}$\,\msun\ comprises of molecular gas, which implies a smaller stellar mass of $M_*\lesssim(0.7-2.1)\times10^{10}$\,\msun. However, the large uncertainties in the derived molecular gas mass and in the dynamical mass prevents us from putting tight constraints in the stellar mass in these quasar hosts.

It should be noted that the uncertainties quoted in this section only include the uncertainties in the measured high-$J$ CO line fluxes and
not the uncertainty in the shape of the CO excitation ladder. For example, if we adopt the observed CO SLED of quasars for which high $J$ and low $J$ CO lines have been measured from \citet{car13}, who find luminosity ratios of \lpcovi/$L^\prime_\mathrm{CO(1-0)}\approx0.34$ and \lpcovii/$L^\prime_\mathrm{CO(1-0)}\approx0.38$, then the estimated CO(1--0) luminosities (and molecular gas masses) are 57--95\% higher compared to the luminosities derived using the J1148+5251 CO SLED (see Figure~\ref{fig:lco}). A future measurement of a low-$J$ CO line with the Jansky Very Large Array is essential to obtain a more accurate estimate of the CO(1--0) luminosity in our quasar hosts. 

We can verify our CO-based molecular gas mass by computing the gas mass from the amount of dust in the host galaxies presented in \citet{ven16}. The quasar hosts have estimated dust masses of $M_d(\mathrm{J}0305-3150)=(4.5-24)\times10^8$\,\msun, $M_d(\mathrm{J}0109-3047)=(0.7-4.9)\times10^8$\,\msun, and $M_d(\mathrm{J}2348-3054)=(2.7-15)\times10^8$\,\msun, which we can use to derive gas masses by assuming the local gas-to-dust mass ratio of 70--100 \citep[e.g.,][]{dra07,san13}. Similar gas-to-dust mass ratios of $\sim$70 have also been found in starburst systems at high redshift \citep[e.g.,][]{rie13,wan16}. Using the local gas-to-dust mass ratio, we obtain (atomic and molecular) gas masses of $(3.2-24)\times10^{10}$\,\msun, $(0.5-4.9)\times10^{10}$\,\msun, and $(1.9-15)\times10^{10}$\,\msun, for J0305--3150, J0109--3047, and J2348--3054, respectively. If we further assume that $\sim$75\% of the dust-derived gas mass is molecular \citep[e.g.,][]{rie13,wan16}, then for J0305--3150 and J0109--3047 the lower values of the dust-derived gas mass agree with the CO-based molecular gas mass. For J2348--3054 the dust mass predicts a higher gas mass.

\subsection{Atomic Carbon Mass}
\label{sec:cimass}

We can make use of the \ci\ line luminosity (or limits thereof) to calculate the mass of atomic carbon in the quasar host galaxies. If we assume that the \ci\ emission is optically thin, then we can use the relation between \ci\ brightness and the mass in neutral carbon given by \citet{wei03,wei05}:
 
\begin{equation}
M_\mathrm{CI}/M_\sun= 4.566\times10^{-4} Q(T_\mathrm{ex})\frac{1}{5}e^{T_2/T_\mathrm{ex}} L^\prime_\mathrm{[CI]},
\end{equation}

\noindent
where $Q(T_\mathrm{ex})=1+3e^{-T_1/T_\mathrm{ex}}+5e^{-T_2/T_\mathrm{ex}}$ is the \ion{C}{1} partition function and $T_1=23.6$\,K and $T_2=62.5$\,K are the energies above the ground state. Following the literature, we set the excitation temperature to $T_\mathrm{ex}=30$\,K \citep[see, e.g.][]{wal11}. Note that if we assume a higher excitation temperature of $T_\mathrm{ex}=50$\,K, the derived $M_\mathrm{CI}$ would be $\sim$38\% lower. 

For the host galaxy of J2348--3054, in which we tentatively detect the \ci\ emission line, we derive a neutral carbon mass of $M_\mathrm{CI}=(1.0\pm0.4)\times10^7$\,\msun. The atomic Carbon abundance, relative to H$_2$ is given by X\ci\,=\,$M_\mathrm{CI}/(6 M_{\mathrm{H}_2}$). For a sample of $z=2-3$, FIR-bright sources, \citet{wal11} derived a carbon abundance of X\ci\,$=(8.4\pm3.5)\times10^{-5}$. Applying this abundance we obtain an independent molecular gas mass in J2348--3054 of $M_{\mathrm{H}_2}=(1.4-3.4)\times10^{10}$\,\msun, which is consistent with the molecular gas mass of $(1.2\pm0.2)\times10^{10}$\,\msun\ derived from the CO observations in Section~\ref{sec:comass}. 

For the other two sources, J0109--3047 and J0305--3150, we can only obtain 3$\sigma$ upper limits of $M_\mathrm{CI}<9.6\times10^6$\,\msun\ and $M_\mathrm{CI}<2.1\times10^7$\,\msun\ (Table~\ref{tab:obsdesc}). The upper limits we obtain for the molecular gas mass assuming the 
neutral carbon abundance from \citet{wal11} of $M_{\mathrm{H}_2}<3\times10^{10}$\,\msun\ (J0109--3047) and $M_{\mathrm{H}_2}<7\times10^{10}$\,\msun\ (J0305--3150) agree well with our molecular gas masses inferred from the CO luminosities.

\subsection{The Characteristics of the ISM}
\label{sec:ism}

With the detection of various FIR lines in our quasar host galaxies, we can start to constrain the physical properties of the ISM in these high redshift galaxies by comparing the luminosity of the emission lines with each other and with the continuum. The observed lines can arise due to star formation in photodissociation regions (PDRs) where the radiation field is dominated by UV photons from young stars, or in X--ray dominated regions (XDRs), where the X--ray radiation from the accreting black hole dominates the emission (or a combination thereof). Alternatively, a substantial fraction of the \cii\ emission could be associated with the diffuse ionized medium. However, observations of local starburst galaxies suggest that only up to 30\% of the \cii\ emission could be emitted by the diffuse ionized medium \citep[e.g.,][]{car94,lor96,col99}. Similarly, studies of high redshift, FIR luminous sources have concluded that, based on high observed \cii/[\ion{N}{2}] line ratios, at most a small fraction of the \cii\ emission comes from the ionized phase of the gas \citep[e.g.,][]{dec14,pav16}. In the remainder of this section, we will assume that the majority of the \cii\ emission comes from the same region as the CO and \ci\ emission.

\begin{figure}
\plotone{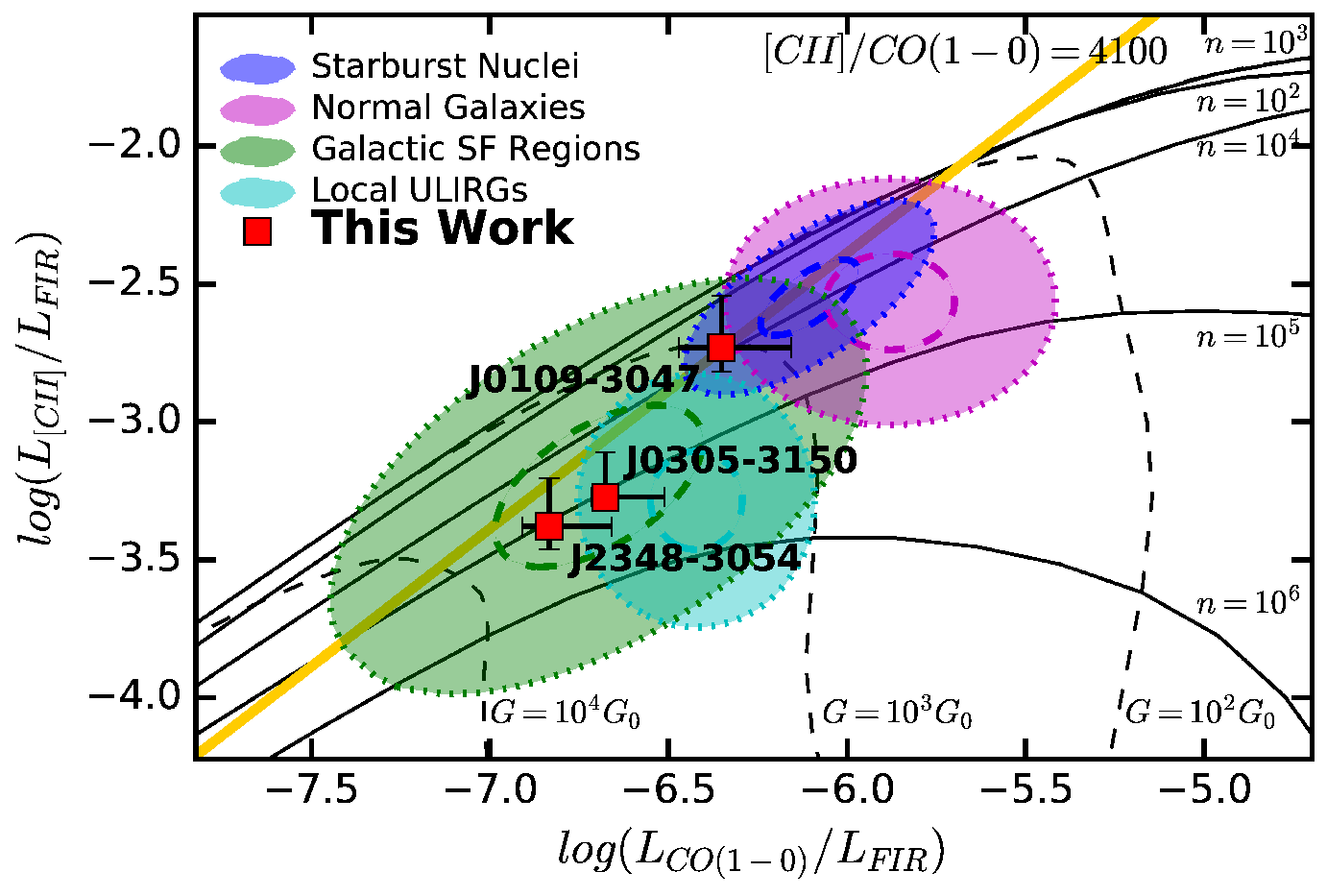}
\caption{\lcii/\lfir\ as function of $L_\mathrm{CO(1-0)}$/\lfir, adapted from \citet{fer14}. Overplotted are values for the UV radiation field $G_0$ and gas density $n$ from the PDR model of \citet{kau06}. The regions in \lcii/\lfir\ and $L_\mathrm{CO(1-0)}$/\lfir occupied by starburst nuclei, normal galaxies, galactic star-forming regions and local ULIRGs are outlined with colored ellipticals. The values derived for the three $z>6.6$ quasar hosts are plotted with  red squares.}
\label{fig:staceyplot}
\end{figure}

An initial diagnostic plot to test the nature of the emission in our sources is shown in Figure~\ref{fig:staceyplot}, where we plot the luminosity ratio \cii/FIR as function of CO(1--0)/FIR. Regions occupied by different galaxies in the nearby universe are shown as well as the typical \cii/CO(1--0) ratio of \lcii/$L_\mathrm{CO(1-0)}=4100$ (Section~\ref{sec:comass}). Additionally, contours for UV field strength and gas density are shown for PDR models from \citet{kau99}. These contours show the phase space where PDRs can describe the observed emission line ratios. Sources falling to the upper left, above the typical \cii/CO(1--0) ratio of 4100, would require a non-PDR emission mechanism such as XDRs to explain the observed lines and continuum. All three of our quasar host galaxies fall in the phase space that can be described by PDRs and are similar to local starbursts and ULIRGs. 

\begin{figure*}
\plottwo{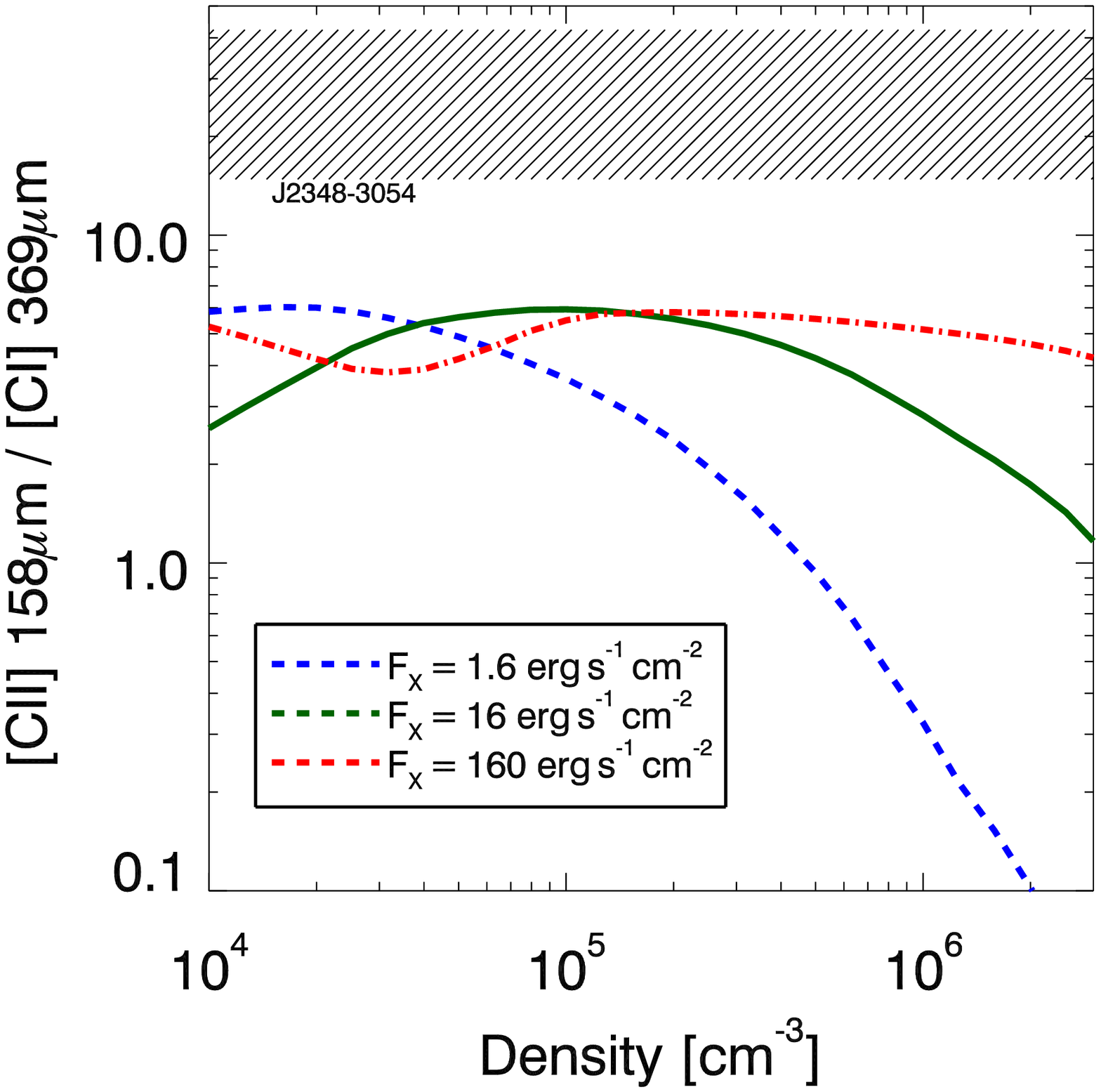}{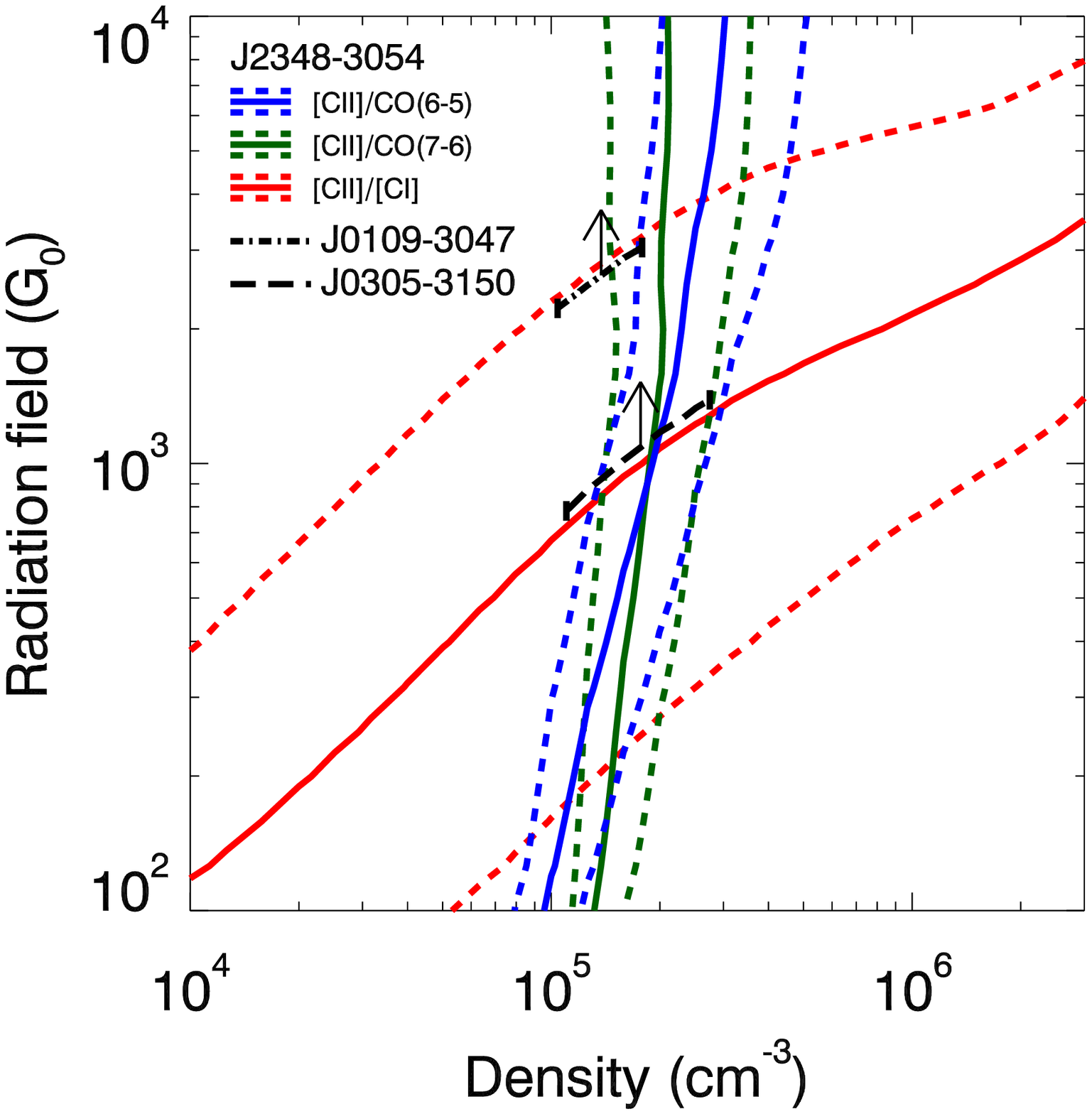}
\caption{Emission line ratio predictions from the XDR (left) and PDR (right) models of \citet{mei07}. On the left, we plot the \cii/\ci\ luminosity ratio as function of density for three different values of the X-ray flux. The \cii/\ci\ line ratio never exceeds a value of about 6 in the XDR models. Based on (the limits on) the line ratio of 24$_{-9}^{+19}$ (hashed region), $>$32, and $>$24 we measure for J2348--3054, J0109--3047, and J0305--3150, respectively, we can exclude that the radiation field in these quasar host galaxies is dominated by the X--ray radiation of the AGN. On the right, we show the constraints on the density and the UV radiation field in the three quasar host galaxies. The red, blue, and green lines show the \cii/\ci, \cii/CO(6--5), and \cii/CO(7--6) line ratio measured in J2348--3054 with the dashed lines indicating the 1$\sigma$ uncertainty in the line ratios. The dotted-dashed and long-dashed lines show the range of densities and the lower limit on the radiation field for J0109--3047 and J0305--3150, respectively.}
\label{fig:pdrxdr}
\end{figure*}

The results from the diagnostic shown in Figure~\ref{fig:staceyplot} can be confirmed, and ISM properties can be derived by applying the PDR and XDR models from
\citet{mei05} and \citet{mei07}. In these models the \cii/CO(7--6) and \cii/CO(6--5) line ratios range from $<$1 to $>$300, depending on the density and radiation field (UV radiation in case of a PDR and X--ray radiation for XDR models). In contrast, the \cii/\ci\ line ratio is predicted to be significantly different for a PDR and an XDR. While for a PDR the \cii/\ci\ ratio is anticipated to lie between $\sim$3--50, the range is predicted to be $\lesssim$6 for XDRs, see Figure~\ref{fig:pdrxdr}. We can compare this range with the values measured in our quasar host galaxies. For J2348--3054, we derive a ratio of \cii/\ci\,$\approx24$, while for J0109--3047 and J0305--3150, the non-detection of the \ci\ line results in lower limits of $>$32 and $>$24, respectively. All these values are significantly higher than the maximum line ratio predicted for X-ray dominated regions in the models of \citet{mei07}. Based on these models we conclude that the heating of the ISM is dominated by star-formation. 

In contrast to the results shown in Figure~\ref{fig:staceyplot}, for J2348--3054 (for which we measured the strength of the \cii\, CO(6--5), CO(7--6), and \ci\ lines) we can constrain the density and radiation field, without relying on assumptions on the shape of the FIR emission and the CO excitation ladder. The results are shown in Figure~\ref{fig:pdrxdr}. From the various line ratios, we derive a density of $\sim$$2\times10^5$\,cm$^{-3}$ and a radiation field strength of $\sim$$10^3$\,$G_0$, with $G_0$ the radiation field strength in Habing units, $1.6\times10^3$\,erg\,s$^{-1}$\,cm$^{-2}$. Also shown in Figure~\ref{fig:pdrxdr} are the constraints on the ISM properties in J0109--3054 and J0305--3150. From the \cii\ to CO line ratios we obtain densities of $n_\mathrm{J0109-3047}\approx10^5$\,cm$^{-3}$ and $n_\mathrm{J0305-3150}\approx2\times10^5$\,cm$^{-3}$. The non-detection of the \ci\ line results in a lower limit on the radiation field strength of $G_\mathrm{J0109-3047}>2\times10^3$\,$G_0$ and $G_\mathrm{J0305-3150}>10^3$\,$G_0$.

If we assume that the \cii\ flux coming from the PDR is only a factor 0.7 of the observed value (as found in local starbursts, see the discussion above), then from the PDR models we derive for J2348--3054 a slightly higher density of $2.3\times10^5$\,cm$^{-3}$ and a lower radiation field strength of $5\times10^2$\,$G_0$. To further constrain the properties of the ISM in these quasar host galaxies, observations of additional FIR emission lines are required.  

With the derived values for the density and the radiation field from the PDR models, we can compare the properties of the ISM in the VIKING quasar hosts with those in other $z>6$ sources. The host galaxies of the quasar J1148+5251 at $z=6.42$ has been detected in many emission lines, including \cii\ \citep[e.g.,][]{mai05,wal09b}, CO \citep[e.g.,][]{ber03b,wal03,rie13,ste15}, and \ci\ \citep{rie09}. The \cii/CO(7--6) ratio is very similar to that of the VIKING quasar hosts, suggesting that the ISM density is also around $2\times10^5$\,cm$^{-3}$. On the other hand, the \cii/\ci\ ratio in J1148+5251, \lcii/$L_\mathrm{[CI]}\approx42$ \citep{rie13}, is higher than we measured in J2348--3054. Based on the PDR models of \citep[][Figure~\ref{fig:pdrxdr}]{mei05}, this would require a radiation field strength of a $\sim$$4\times10^3$\,$G_0$. \citet{wan16} found very comparable values for the density and radiation field in the host of quasar J0100+2802 at $z=6.3$ based on the strength of the \cii\, CO(6--5), and CO(2--1) emission lines and the FIR luminosity: $n=1\times10^5$\,cm$^{-3}$ and $G=4\times10^3$\,$G_0$. Although both the quasar luminosity and the derived values for $G$ are higher for J1148+5251 and J0100+2802 compared to the VIKING quasars and may point to a relation between the quasar luminosity and $G$, there is no evidence for such a relation among our VIKING sources: the faintest quasar in our sample is J0109--3047, both in quasar luminosity \citep{ven13} and in FIR luminosity \citep{ven16}, and has the highest (limit on) $G$. This is confirmed when analyzing the emission line ratios in the starburst HFLS3 at $z=6.34$ \citep{rie13}, a very massive galaxy without a luminous AGN. Again, the \cii/CO(7--6) ratio in HFLS3 is very similar to those in the $z>6.5$ quasar hosts, while the \cii/\ci\ line ratio is $>$51 which implies a very high radiation field (Figure~\ref{fig:pdrxdr}).

\section{SUMMARY}
\label{sec:summary}

In this paper, we present ALMA band 3 observations of the CO(6--5), CO(7--6), and \ci\,369\,$\mu$m emission lines in three of the highest redshift quasar host galaxies at $6.6<z<6.9$. The sample has been previously detected in \cii\,158\,$\mu$m emission and the underlying FIR dust continuum \citep{ven16}. CO is detected at high significance in all sources, making these measurements the highest-redshift CO detections to date. Given the resolution of our observations ($\gtrsim$2\farcs5, or $>$12\,kpc), all quasar hosts are spatially unresolved in the current data.

In two of our sources, we detect the continuum emission around the CO emission lines (around 400\,$\mu$m rest-frame). A comparison with the previously measured dust continuum at higher frequency (close to the \cii\ line) gives dust temperatures that are broadly consistent with the canonical value of 47\,K, albeit the current uncertainties are very large. Future observations of multi--frequency continuum emission clearly have the potential to derive more accurate dust temperatures, and possibly spatially resolved temperature gradients.

Assuming a CO SLED similar to that observed in the $z=6.4$ quasar J1148+5251, we derive molecular gas reservoirs of $1-3\times10^{10}$\,\msun\ in the quasar hosts. For J2348--3054 and J0109--3047 we estimate high stellar masses of $M_*=(2-17)\times10^{10}$\,\msun. For J0305--3150 as much as $\sim$54--81\% of the dynamical mass is in the form of the molecular gas, indicating that the stellar mass is $M_*<2.1\times10^{10}$\,\msun, less than 22$\times$ the mass of the central black hole \citep[$M_\mathrm{BH}=9.5\times10^8$\,\msun;][]{der14}. Possibly we overestimated the molecular gas mass in this quasar host (if, for example, the CO(6--5)/CO(1--0) and CO(7--6)/CO(1--0) luminosity ratios are larger in J0305--3150 than that measured in J1148+5251), or the dynamical mass is significantly larger than the mass traced by the \cii\ emission. 
The extrapolated \cii\--to--CO(1--0) luminosity ratio is 2500--4200, consistent with measurements in galaxies at lower redshift. 

The (marginal) detection of \ci\ emission in one quasar host galaxy (at $z=6.9$) and the limit on the \ci\ emission in the other two galaxies enables us to characterize the physical properties of the interstellar medium in $z\sim7$ quasar hosts for the first time. In this case, the derived global CO/\cii/\ci\ line ratios are consistent with them emerging from photodissociation regions (PDRs), but inconsistent with an excitation from X--ray dominated regions (XDR). This implies that if the central supermassive black hole gives rise to an XDR, it does not dominate the excitation of the overall gas reservoir. This finding provides further evidence that the quasar host galaxies studied in this paper harbor intense starbursts, and provides justification to link quantities derived for the molecular gas and dust content to ongoing star formation in these quasar hosts.  

The observations presented here represent only a modest investment of ALMA time (with typical on-source integration times between 15 and 35 minutes). This implies that future observations have the potential to resolve the molecular gas emission, which eventually could lead to spatially resolved excitation measurements within the quasar host galaxies at the highest redshift. Furthermore, by targeting other FIR emission lines, such as [\ion{O}{1}] 146\,$\mu$m, [\ion{N}{2}] 122\,$\mu$m, and [\ion{O}{3}] 88\,$\mu$m, we will be able to put additional constraints on the properties and metallicity of the ISM in these forming massive galaxies in the early Universe.

\acknowledgments We thank the referee for providing good comments that helped to improve the manuscript. B.P.V.\ and F.W.\ acknowledge funding through the ERC
grant ``Cosmic Dawn''. Support for R.D.\ was provided by the DFG
priority program 1573 ``The physics of the interstellar medium''. R.G.M.\ acknowledges the support of the UK Science and
Technology Facility Council (STFC). This
paper makes use of the following ALMA data:
ADS/JAO.ALMA\#2013.1.00273.S. ALMA is a partnership of ESO
(representing its member states), NSF (USA) and NINS (Japan), together
with NRC (Canada) and NSC and ASIAA (Taiwan), in cooperation with the
Republic of Chile. The Joint ALMA Observatory is operated by ESO,
AUI/NRAO and NAOJ.

\facility{ALMA}.

\end{document}